**The electron-phonon coupling constant, the Fermi temperature and unconventional superconductivity in the carbonaceous sulfur hydride 190 K superconductor**

E. F. Talantsev[1,2]

[1]M.N. Miheev Institute of Metal Physics, Ural Branch, Russian Academy of Sciences, 18, S. Kovalevskoy St., Ekaterinburg, 620108, Russia

[2]NANOTECH Centre, Ural Federal University, 19 Mira St., Ekaterinburg, 620002, Russia

E-mail: evgeny.talantsev@imp.uran.ru

## Abstract

Recently, Snider *et al* (2020 *Nature* **586** 373) reported on the observation of superconductivity in highly-compressed carbonaceous sulfur hydride, $H_x(S,C)_y$. The highest critical temperature in $H_x(S,C)_y$ by 5 K exceeds previous record of $T_c$ = 280 K reported by Somayazulu *et al* (2019 *Phys. Rev. Lett.* **122** 027001) for highly-compressed $LaH_{10}$. In this paper we analyse experimental temperature dependent magnetoresistance data, $R(T,B)$, reported by Snider *et al*. The analysis shows that $H_x(S,C)_y$ compound exhibited $T_c$ = 190 K ($P$ = 210 GPa) has the electron-phonon coupling constant $\lambda_{e\text{-}ph}$ = 2.0 and the ratio of critical temperature, $T_c$, to the Fermi temperature, $T_F$, in the range of $0.011 \leq T_c/T_F \leq 0.018$. These deduced values are very close to ones reported for $H_3S$ at $P$ = 155-165 GPa (Drozdov *et al* 2015 *Nature* **525** 73). This means that in all considered scenarios the carbonaceous sulfur hydride 190 K superconductor falls into unconventional superconductors band in the Uemura plot, where all other highly-compressed super-hydride/deuterides are located. It should be noted that that our analysis shows that all raw $R(T,B)$ datasets for $H_x(S,C)_y$ samples for which Snider *et al* (2020 *Nature* **586** 373) reported $T_c$ > 200 K cannot be characterised as reliable data sources. Thus, independent experimental confirmation/disprove high-$T_c$ values in the carbonaceous sulfur hydride is required.

## I. Introduction

In 2015 Drozdov *et al* [1] discovered the first near-room-temperature (NRT) superconductor highly-compressed sulphur hydride, $H_3S$. To date, NRT superconductivity has been observed in four super-hydrides/deuterides systems subjected to high pressure: Th-H [2], S-(H,D) [1,3-5], Y-H [6,7], La-(H,D) [8-10].

Recently, Snider *et al* [11] reported on the observation that ternary compound $H_x(S,C)_y$ exhibits the transition temperature within a range of $T_c$ = 275-287 K (at $P$ = 258-270 GPa), which is by about 5 K higher than previous record of $T_c$ = 280 K, reported by Somayazulu *et al* [8] in highly-compressed $LaH_{10}$. While detailed phase/structural and phonon spectrum and other physical properties measurements [1-7,9,12-14], as well as the first-principle calculation studies [2,6,15-27] are on-going tasks for $H_x(S,C)_y$ compound [28], in this paper we report results of the analysis of temperature dependent magnetoresistance, $R(T,B)$, from which we deduced:

1. The charge carriers effective mass, $m_{eff}$;
2. The ground state superconducting coherence length, $\xi(0)$.
3. The Fermi temperature, $T_F$.

In a result, we find that in all considered scenarios $H_x(S,C)_y$ exhibited $T_c$ = 190 K ($P$ = 210 GPa) has the ratio of $T_c/T_F$ within a range of $0.011 < T_c/T_F < 0.018$. This means that $H_x(S,C)_y$ falls in to unconventional superconductors band in the Uemura plot, where heavy-fermions, cuprates, pnictides and all near-room-temperature superconductors are located.

## II. Description of the approach

Detailed description of the approach can be found elsewhere [29]. In short, we use the $T_c/T_F$ ratio to locate the position of the superconductor in the Uemura plot [30]. $T_F$ is calculated by:

$$T_F = \frac{\pi^2}{8 \cdot k_B} \cdot m_{eff}^* \cdot \xi^2(0) \cdot \left(\frac{\alpha \cdot k_B \cdot T_c}{\hbar}\right)^2, \quad (1)$$

where $\hbar = h/2\pi$ is the reduced Planck constant, $k_B$ is the Boltzmann constant, $\alpha = \frac{2 \cdot \Delta(0)}{k_B \cdot T_c}$, where $\Delta(0)$ is the ground state energy gap. We deduce the ground state coherence length, $\xi(0)$, by the fit of experimental $R(T,B)$ curves to analytical approximate of Werthamer, Helfand and Hohenberg theory [31,32] proposed by Baumgartner *et al* [33]:

$$B_{c2}(T) = \frac{\phi_0}{2 \cdot \pi \cdot \xi^2(0)} \cdot \left(\frac{\left(1-\frac{T}{T_c}\right) - 0.153 \cdot \left(1-\frac{T}{T_c}\right)^2 - 0.152 \cdot \left(1-\frac{T}{T_c}\right)^4}{0.693}\right). \quad (2)$$

It should be noted that we define $T_c$ and $B_{c2}$ by employing the most strict criterion, i.e. $\frac{R(T)}{R_{norm}} \to 0$ (detailed discussion of the problem can be found elsewhere [34]). We will designate this model as B-WHH model.

Because $\alpha = \frac{2 \cdot \Delta(0)}{k_B \cdot T_c}$ cannot be deduced from available experimental data, we perform our calculations by assuming that $\alpha$ has lower and upper limits within *s*-wave superconducting gap symmetry:

$$3.5 \leq \alpha \leq 4.5 \quad (3)$$

where the lower limit is the weak-coupling limit of Bardeen-Cooper-Schrieffer theory [35] and the upper limit is the value for anharmonic phonons computed for precursor compound $H_3S$ by first principle calculations approach [19,36].

The charge carriers effective mass, $m_{eff}^*$, has been calculated by the use of the Eliashberg's theory expression [37]:

$$m_{eff}^* = \left(1 + \lambda_{e-ph}\right) \cdot m_e \quad (4)$$

where $m_e$ is the electron mass and the electron-phonon coupling constant $\lambda_{e-ph}$ is deduced by the use of the advanced McMillan equations [34,38]:

$$T_c = \left(\frac{1}{1.45}\right) \cdot T_\theta \cdot e^{-\left(\frac{1.04 \cdot \left(1+\lambda_{e-ph}\right)}{\lambda_{e-ph} - \mu^* \cdot \left(1+0.62 \cdot \lambda_{e-ph}\right)}\right)} \cdot f_1 \cdot f_2^* \quad (5)$$

where $T_\theta$ is the Debye temperature, and

$$f_1 = \left(1 + \left(\frac{\lambda_{e-ph}}{2.46\cdot(1+3.8\cdot\mu^*)}\right)^{3/2}\right)^{1/3} \quad (6)$$

$$f_2^* = 1 + (0.0241 - 0.0735\cdot\mu^*)\cdot\lambda_{e-ph}^2. \quad (7)$$

where $\mu^*$ is the Coulomb pseudopotential parameter (ranging from $\mu^*$ = 0.13 0-0.16 [19,34,38]) for which in this paper we use an average value of $\mu^*$ = 0.13. And the Debye temperature was deduced from the fit of $R(T, B = 0)$ data to Bloch-Grüneisen (BG) equation [39,40]:

$$R(T, B = 0) = R_0 + A\cdot\left(\frac{T}{T_\theta}\right)^5\cdot\int_0^{\frac{T_\theta}{T}}\frac{x^5}{(e^x-1)\cdot(1-e^{-x})}\cdot dx \quad (8)$$

where $T_\theta$, $R_0$ and $A$ is a free-fitting parameter. It should be noted, that the procedure to deduce the electron-phonon coupling constant, $\lambda_{e-ph}$, by combined use of the Bloch-Grüneisen and the McMillan equations is widely used in the field [41,42].

It is important to note that analysed $R(T,B)$ datasets for $H_x(S,C)_y$ were directly extracted from Figs. 1,2 of Ref. 11 (because published $R(T,B)$ datasets are not available from the authors of Ref. 11, because these published $R(T,B)$ curves will be in use in a patent application [43]). Thus, we extract $R(T,B)$ datasets for the $H_x(S,C)_y$ samples by the use of the same routine as it has been done for dozens of $R(T,B)$ datasets for other highly-compressed superconductors, which we analysed in our recent papers [29,34]. For instance, we can mention: black phosphorous (raw data reported by Shirotani *et al* [44] in their figure 5), boron (raw data reported by Eremets *et al* [45] in their figure 2), germanium arsenide (raw data reported by Liu *et al* [46] in their figure 3), silane (raw data reported by Eremets *et al* [47] in their figure 2), ζ-phase of $O_2$ (raw data reported by Shimizu *et al* [48,49,50]), sulphur (raw data reported by Shimizu *et al* [51] in their Fig. 10), lithium (raw data reported by Shimizu *et al* [52] in their Fig. 2), sulphur hydride/deuteride (raw data reported by Einaga *et al* [53] in their Fig. 3(a,b); by Drozdov *et al* [1] in their figure 2(b); and by Mozaffari *et al* [54] in their figure 1), lanthanum hydride/deuteride (raw data reported by Drozdov *et al* [55] in their figures 1,2,4, and extended data figures 2,3,5).

## III. Results and discussion

Snider *et al* [11] in their Fig. 1 reported several $R(T,B = 0)$ curves for three highly-compressed $H_x(S,C)_y$ samples. To be reliably fitted to Eq. 8, the $R(T,B = 0)$ curve should be measured at reasonably wide temperature range and also should not to be distorted by any experimental artefact (for instance, the failure of electronics or diamond anvil cell, or both), which change the shape of $R(T,B)$ curve, and more likely observed drop in resistance, which can be mistakenly interpreted as the superconducting transition. We find, that from 9 presented in Figs. 1,2 [11] experimental $R(T,B=0)$ curves, the only one dataset designated as "Run 3" at $P$ = 210 GPa (Fig. 1) can, with satisfaction quality, fit to Eq. 8. It should be noted that fits of other datasets to Eq. 8 either do not converge, either after the converging, deduced values have some large uncertainty boundaries, exceeding deduced values. This is a very unusual behaviour of $R(T,B=0)$ datasets for $H_x(S,C)_y$ samples reported by Snider *et al* [11], because 27 $R(T,B=0)$ datasets for a variety of highly-compressed superconductors (ranging from elements to superhydrides/superdeuterides) reported by different research groups [1,44-55] were fitted to Eq. 8 with a good quality and accuracy [29,34]. In this paper, we analyse raw $R(T,B)$ dataset for $H_x(S,C)_y$ sample ("Run 3" at $P$ = 210 GPa (Fig. 1) [11]) in an assumption that this dataset is accurate and correct. However, it should be stressed that independent confirmation/disprove primary results reported by Snider *et al* [11], including very high-$T_c$ values, is required [56].

In Fig. 1 we show the fit of $R(T,B = 0)$ curve (Run 3, $P$ = 210 GPa) to Eq. 8. Deduced Debye temperature is $T_\theta = 1497 \pm 8\ K$. By the use of Eqs. 5-7, and taking in account $T_c$ = 190 K, one can calculate $\lambda_{e-ph} = 2.0$. This value is in a very good agreement with computed $\lambda_{e-ph} = 1.84\ (200\ GPa)$ and $\lambda_{e-ph} = 1.71\ (250\ GPa)$ reported by Errea *et al* [17] for precursor binary compound $H_3S$. Based on pivotal result of Duan *et al* [54] who calculated $\lambda_{e-ph} = 2.19\ (200\ GPa)$ for $H_3S$, we can conclude that deduced by us $\lambda_{e-ph} = 2.0$ for $H_x(S,C)_y$ (Run 3, $P$ = 210 GPa) is very close to $\lambda_{e-ph}$ for binary $H_3S$ compound. This results may allude that sample "Run 3" at $P$ = 210 GPa (Fig. 1) [11] can be just binary $H_3S$ compound. Based on deduced $\lambda_{e-ph}$ value, the charge carriers effective mass is:

$$m^*_{eff} = (1 + \lambda_{e-ph}) \cdot m_e = 3.0 \cdot m_e \qquad (9)$$

It should be noted that this value is in a good agreement with $m^*_{eff} = 2.76 \cdot m_e$ reported by Durajski [16] for the precursor compound $H_3S$.

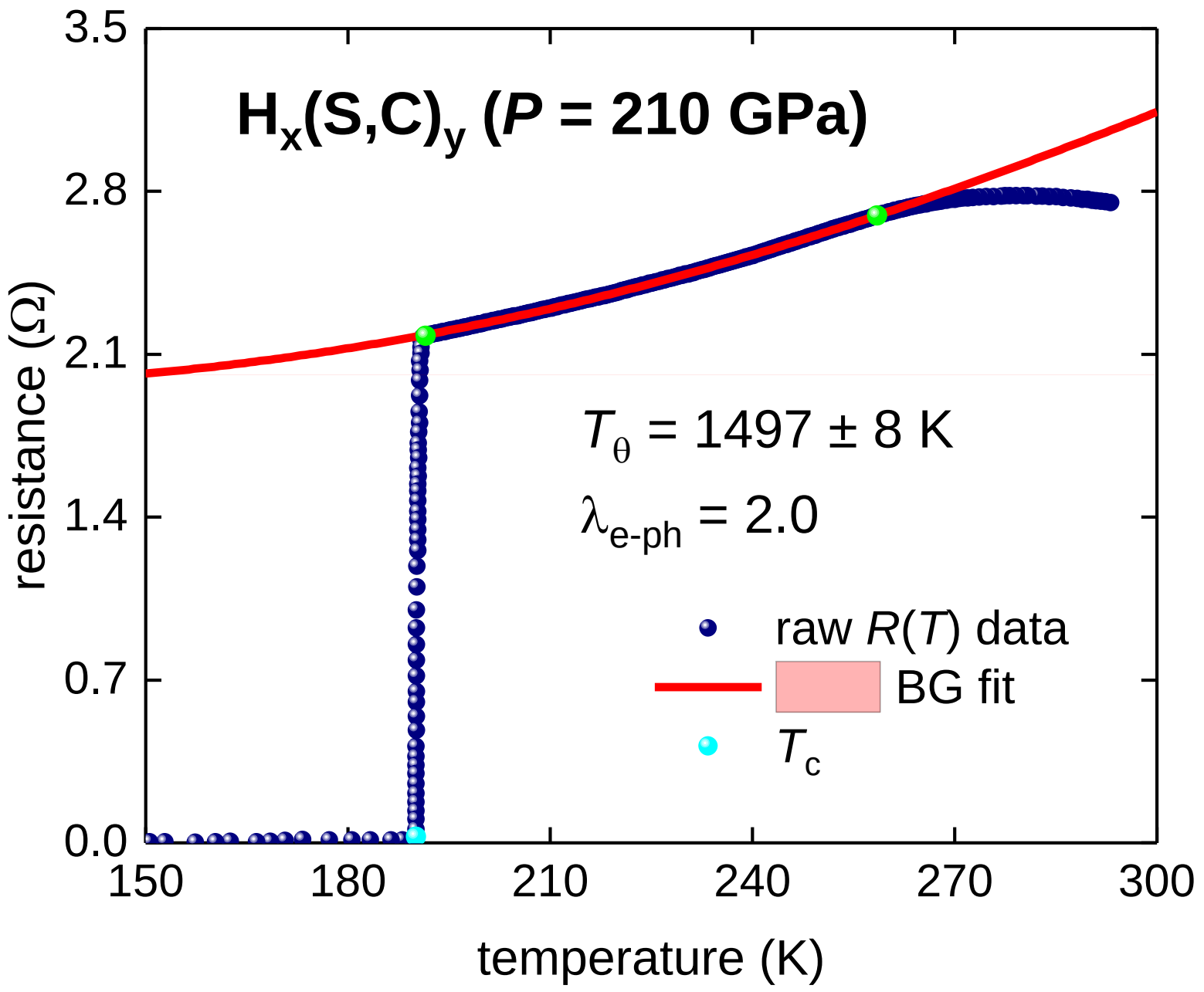

**Figure 1.** Resistance data, $R(T)$, and fit to BG model (Eq. 8) for $H_x(S,C)_y$ sample compressed at $P$ = 210 GPa (raw data is from Ref. 11, where the sample designated as *Run 3*). The fit quality is $R$ = 0.9998. 95% confidence bar is shown. Green balls show bounds for which $R(T)$ data was used for the Eq. 8 fit.

Snider *et al* [11] in their Fig. 2(b) reported $B_{c2}(T)$ data for this sample. We extract $B_{c2}(T)$ data directly from Fig. 2(b) of Ref. 11. The fit of $B_{c2}(T)$ to Eq. 2 is shown in Fig. 2(a). Deduced ground state coherence length is $\xi(0) = 2.39 \pm 0.04\ nm$. It should be noted that Eq. 2 is a good extrapolative tool, that can be proved for the case of highly-compressed $LaH_{10}$. Drozdov *et al.* [9] reported first $B_{c2}(T)$ dataset for this NRT superconductor for applied field up to $B_{appl}$ = 9 T. This dataset was used to extrapolate $B_{c2}(0)$ value for $LaH_{10}$ in Ref. [57]. Recently, Sun et al [58] report new experimental $B_{c2}(T)$ dataset for $LaH_{10}$, which was measured on the world-top magnetic field facilities with applied field up to $B_{appl}$ = 60 T. It can be seen that new experimental data is pretty much reproduced extrapolated values obtained by the use of Eq. 2 for dataset measured up to $B_{appl}$ = 9 T in Ref. 9.

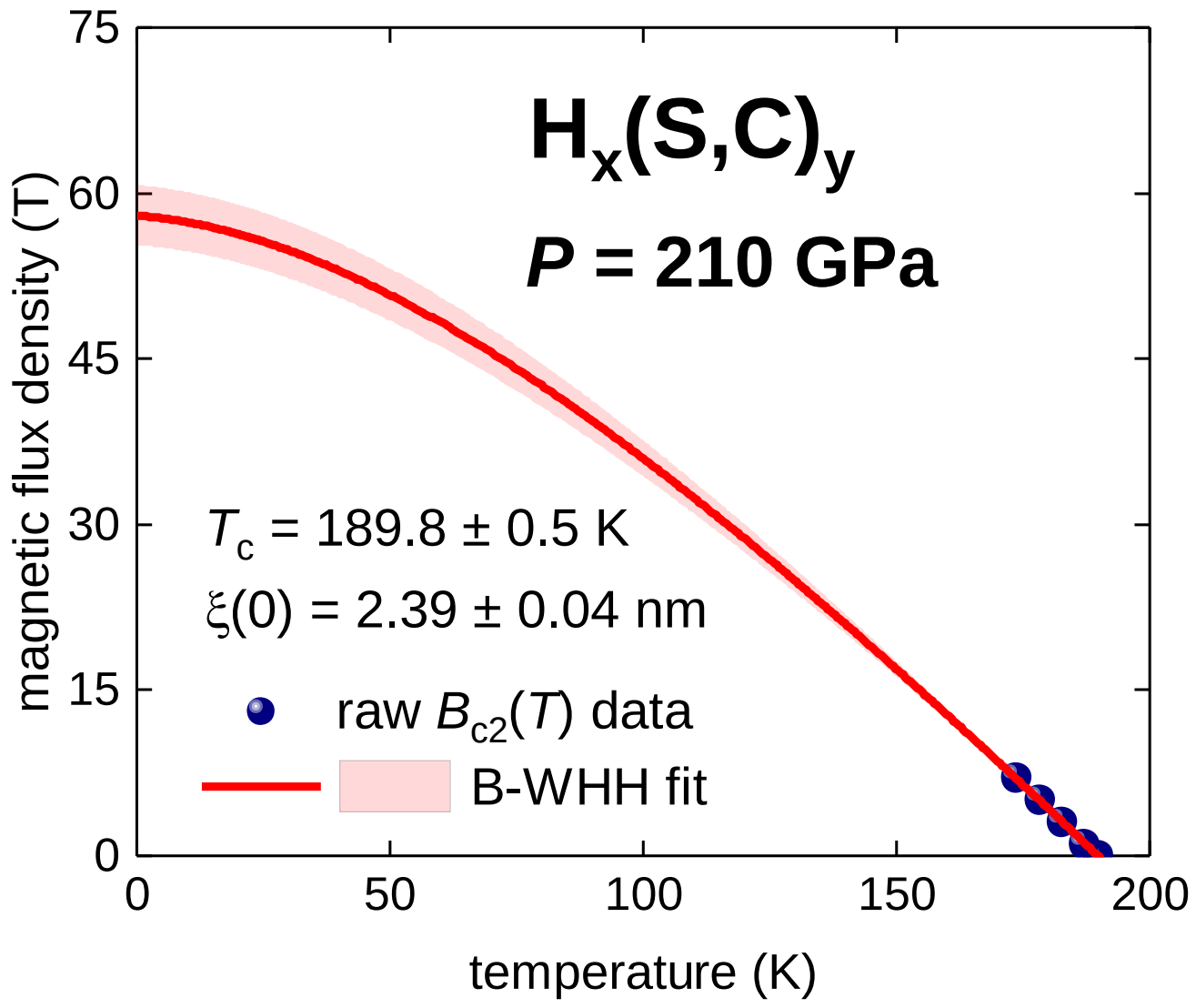

**Figure 2.** Superconducting upper critical field, $B_{c2}(T)$, and data fit to B-WWH model (Eq. 2) for two $H_x(S,C)_y$ samples compressed at $P$ = 210 GPa (raw data is from Ref. 11). The fit quality is $R$ = 0.9993. 95% confidence bars are shown.

From deduced ξ(0), $m^*_{eff}$, and $T_c$, we calculate value ranges for $T_F$ and $T_c/T_F$, by assuming α within its upper and lower limits (Eq. 3). The results are in Table 1.

**Table I.** Deduced and calculated parameters for carbonaceous sulfur hydride. For calculations we use deduced $m^*_{eff} = 3.0 \cdot m_e$.

| *Pressure* (GPa) | Deduced $T_c$ (K) | Deduced ξ(0) (nm) | Assumed $\frac{2\cdot\Delta(0)}{k_B \cdot T_c}$ | $T_F$ ($10^4$ K) | $T_c/T_F$ |
|---|---|---|---|---|---|
| 210 | 189.8 ± 0.5 | 2.39 ± 0.04 | 3.5 | 1.05 ± 0.03 | 0.018 ± 0.001 |
| | | | 4.5 | 1.74 ± 0.06 | 0.011 ± 0.001 |

Analysed sample of highly-compressed $H_x(S,C)_y$ are shown in the Uemura plot (Fig. 3). It can be seen, that in all considered scenarios the carbonaceous sulfur hydride has $0.011 \leq T_c/T_F \leq 0.018$ and falls in the unconventional superconductors band, where heavy-fermions, pnictides, cuprates and highly-compressed near-room-temperature superconductors are located.

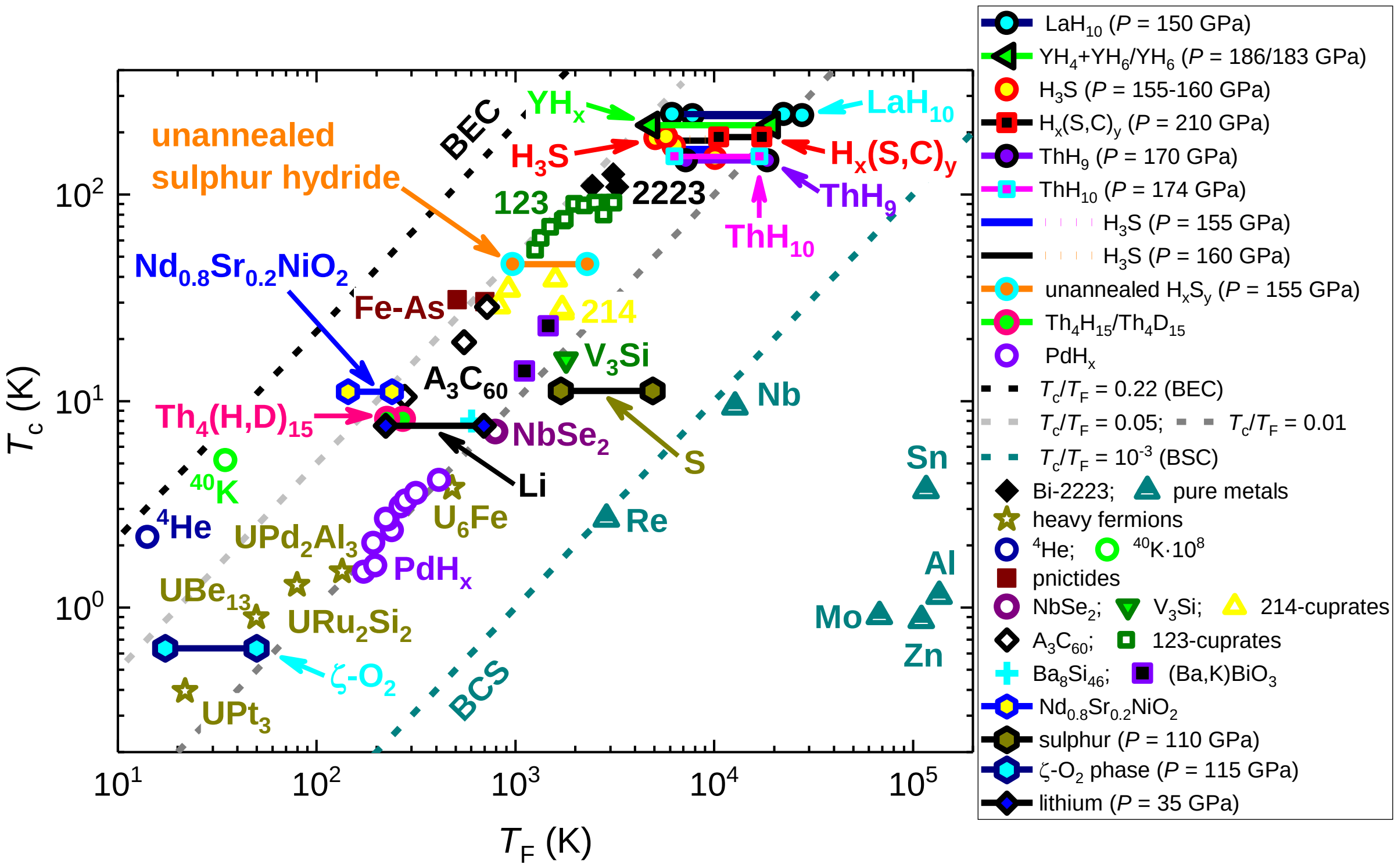

**Figure 3.** A plot of $T_c$ versus $T_F$ where $H_x(S,C)_y$ sample (red/black squares) is shown together with main superconducting families: elemental superconductors, heavy-fermions, pnictides, cuprates, and near-room-temperature superconductors. References on raw data can be found in Refs. 29,30,58,60-63. Boundary lines for the Bose-Einstein condensate (BEC), the Bardeen-Cooper-Schrieffer (BCS) superconductors, and for ratio of $T_c/T_F$ = 0.05, 0.01 are shown for clarity.

There is also a need to clarify that the London penetration depth, $\lambda(T)$, for which Snider *et al* [11] calculate the value of $\lambda(0)$ = 2 nm, cannot be derived in a way presented in Extended Data Fig. 3,b [11]. The estimation for $\lambda(0)$ can be made by taking in account that practically all non-elemental superconductors are type-II superconductors, and most of them have the Ginsburg-Landau parameter $\kappa(0) = \frac{\lambda(0)}{\xi(0)}$ within a range of $60 < \kappa(0) < 120$ [64-71]. Based on deduced $\xi(0)$ = 2.4 nm, expected value for $\lambda(0)$ is within a range: 144 nm < $\lambda(0)$ < 290 nm.It should be noted that binary highly-compressed $H_3S$ has $\lambda(0)$ = 188 nm [71].

The amplitude of the ground state superconducting gap, $\Delta(0)$, also cannot be calculated in a way, as proposed by Snider *et al* [11]. The estimate can be obtained by the use of Eq. 3, from which one can obtain: 29 meV < $\Delta(0)$ < 37 meV It should be noted that temperature and field dependences of superconducting gap, $\Delta(T,B)$, cannot be calculated by the equation:

$$\Delta(T) = \Delta(0) \cdot 1.76 \cdot k_B \cdot T_c \cdot \sqrt{\left(1 - \frac{T}{T_c}\right)}, \quad (10)$$

which was used by Snider *et al* [11] in their Extended Data Fig. 3,d. An accurate analytical expression for $\Delta(T,B=0)$ for *s*-wave symmetry has been given by Gross-Alltag *et al* [72]:

$$\Delta(T) = \Delta(0) \cdot \tanh\left[\frac{\pi \cdot k_B \cdot T_c}{\Delta(0)} \cdot \sqrt{\eta \cdot \frac{\Delta C}{C} \cdot \left(\frac{T_c}{T} - 1\right)}\right] = \Delta(0) \cdot \tanh\left[\frac{2 \cdot \pi}{\alpha} \cdot \sqrt{\eta \cdot \frac{\Delta C}{C} \cdot \left(\frac{T_c}{T} - 1\right)}\right], \quad (11)$$

where $\Delta C/C$ is the relative jump in electronic specific heat at $T_c$, and $\eta$ = 2/3.

## IV. Conclusions

The discovery of superconductivity in highly-compressed $H_3S$ by Drozdov *et al* [1] heralded the era of room-temperature superconductivity. To date, several superhydride/superdeuteride superconductors have been synthesised: $BaH_{12}$ [25], $PrH_9$ [73], $ThH_9$/$ThH_{10}$ [2], $YH_4$/$YH_6$ [6,7], $LaH_{10}$/$LaD_{11}$ [8,9,14], $H_x(S,C)_y$ [11]. The latter compound, in accordance with recent report by Snider *et al* [11], exhibits $T_c$ = 280 K at pressure of $P$ = 267 GPa, which is 5 K higher than $T_c$ = 280 K in $LaH_{10}$ ($P$ = 195 GPa) reported by Somayazulu *et al* [8]. In this paper we extract data reported in Figs. 1,2 of Ref. 11 and deduce the electron-phonon coupling constant, $\lambda_{e\text{-}ph}$ = 2.0, the ground state coherence length, $\xi(0)$ = 2.20 ± 0.09 nm, and the Fermi temperature, $T_F$ = (1.7-2.8) $10^4$ K, for the $H_x(S,C)_y$ superconductor with $T_c$ = 190 K.  Deduced values and reported $T_c$ are indistinguishably close to values reported previously for primary binary compound of $H_3S$. From deduced values, we calculate the $T_c/T_F$ ratio in $H_x(S,C)_y$ compound ($T_c$ = 190 K) and find that in all considered scenarios this compound should be classified as unconventional superconductors.

We should stress that our analysis reveals significant problems with all reported by Snider *et al* [11] $R(T,B)$ data for which claimed $T_c$ > 200 K. These findings are in a good accord with results of independent analysis of $R(T,B)$ data reported by Hirsch and Marsiglio [56]. Based on this, independent confirmation/disprove high-$T_c$ values in highly-compressed $H_x(S,C)_y$ is required.

**Acknowledgement**

Author thanks financial support provided by the state assignment of Minobrnauki of Russia (theme “Pressure” No. AAAA-A18-118020190104-3) and by Act 211 Government of the Russian Federation, contract No. 02.A03.21.0006.